\documentclass[twocolumn,prd,showpacs]{revtex4}
\usepackage{graphicx}

\newcommand{\be}{\begin{equation}}
\newcommand{\ee}{\end{equation}}
\newcommand{\bea}{\begin{eqnarray}}
\newcommand{\eea}{\end{eqnarray}}
\newcommand{\nn}{\nonumber}
\newcommand{\slsh}[1]{{\not \! #1} }
\usepackage{color}
\begin{document}

\title{Confinement in Maxwell-Chern-Simons Planar Quantum Electrodynamics and the $1/N$ approximation}

\author{Christoph P. Hofmann$^1$, Alfredo Raya$^2$, and Sa\'ul S\'anchez Madrigal$^2$}
 \affiliation{$^1$Facultad de Ciencias, Universidad de Colima, Bernal D\'{\i}az del Castillo 340, Colima, Colima 28045, M\'exico.\\
$^2$Instituto de F\'{\i}sica y Matem\'aticas,
Universidad Michoacana de San Nicol\'as de Hidalgo, \\Edificio C-3, Ciudad Universitaria, Morelia, Michoac\'an 58040, M\'exico.}

\date{\today}

\begin{abstract} 
We study the analytical structure of the fermion propagator in planar quantum electrodynamics coupled to a Chern-Simons term  within a four-component spinor formalism. The dynamical generation of parity-preserving and parity-violating fermion mass terms is considered, through the solution of the corresponding Schwinger-Dyson equation for the fermion propagator at leading order of the $1/N$ approximation in Landau gauge. The  theory undergoes a first order phase transition toward chiral symmetry restoration when the Chern-Simons coefficient $\theta$ reaches a critical value which depends upon the number of fermion families considered. Parity-violating masses, however, are generated for arbitrarily large values of the said coefficient.
On the confinement scenario, complete charge screening  --characteristic of the $1/N$ approximation-- is observed in the entire $(N,\theta)$-plane through the local and global properties of the vector part of the fermion propagator.
\end{abstract}


\maketitle

\section{Introduction}
\label{intro}

Quantum electrodynamics in (2+1)-dimensions (QED$_3$) has been a subject of intense studies for more than two decades. One of the reasons is that it serves as a toy model for QCD, because QED$_3$ exhibits the two crucial features of dynamical symmetry breaking and confinement. A natural extension of this model is to add a Chern-Simons (CS) term for the gauge field which explicitly breaks parity. In particular, QED$_3$ with a CS term has found many applications to planar condensed matter phenomena, where it was proposed to serve as an effective theory for high-$T_c$ superconductivity~\cite{supercon1,supercon2,supercon3}, for the quantum Hall effect and, more recently, for graphene~\cite{graphene1,graphene2,graphene3}.

In pure QED$_3$, i.e., without a CS term, at leading order in the $1/N$ expansion, it was found that there exists a critical number $N_c = 32/\pi^2 \approx 3.24$ of fermion families, above which chiral symmetry is restored~\cite{Appel,Appelquist}. This view, however, has been challenged in Ref.~\cite{Pennington}, where dynamical symmetry breaking is found for arbitrarily large values of $N$, although suppressed exponentially. However, lattice calculations~\cite{hands,kogut}, as well as more recent theoretical studies~\cite{AppTL,fischer}, seem to favor a finite critical value $N_c$. Conditions on the existence of a finite $N_c$ in the theory have been recently considered in Ref.~\cite{BCRR}, exploring the infrared behavior of fermion wavefunction renormalization and photon vacuum polarization near criticality.

It is interesting to ask oneself how these general features of pure QED$_3$ are modified by the introduction of a CS term~\cite{Rao,Appelcs,Matsuyama1,Matsuyama2,Matsuyama3,BRS}. Indeed, this question has been addressed in great detail in Refs.~\cite{KondMarPRL,KondMarPRD}. The authors find that, also in presence of a CS term, there exists a critical number $N_c$, above which chiral symmetry breaking ceases to take place. Remarkably, the authors also find a critical value of the CS coefficient which leads to chiral symmetry restoration. The corresponding phase transition is of first order. The authors derive their results by studying the Schwinger-Dyson equation (SDE) for the fermion propagator in a  momentum dependent  gauge selected to yield trivial wavefunction renormalization, thus ensuring that the Ward-Green-Takahashi identities (WGTI) are satisfied.

In our present paper, we will also make use of the four-dimensional reducible representation of the Euclidean Clifford algebra. However, rather than building an equation to have the desired solution by adopting the non-local gauge advocated in Refs.~\cite{KondMarPRL,KondMarPRD}, we find it convenient to analyze the SDE for the fermion propagator in Landau gauge. As we will demonstrate, in the rainbow-ladder approximation, this gauge is approximately consistent with the WGTI.
Our results are in agreement with the findings of Refs.~\cite{KondMarPRL,KondMarPRD}: in particular,  we also find a
critical value for the CS coefficient above which chiral symmetry is restored and confirm that the corresponding phase transition is of first order. However, by studying in detail the question of confinement in QED$_3$ with a CS term, we go beyond the results of Refs.~\cite{KondMarPRL,KondMarPRD}.

Our paper is organized as follows: In Sec.~\ref{model} we define our four-dimensional reducible representation of (2+1)-dimensional quantum electrodynamics with a CS term. The solutions to the gap equation in the leading $1/N$ approximation  are presented in Sec.~\ref{DynamicalFM}.  The question of confinement in QED$_3$ with a CS term is addressed in Sec.~\ref{confinement}. Finally, in Sec.~\ref{conclusions}, we present our conclusions.

\section{The model}
\label{model}

In this article we consider quantum electrodynamics in (2+1)-dimensions with a Chern-Simons term. For fermions, there exist two irreducible two-dimensional representations of the Euclidean Clifford algebra $\{ \gamma_\mu, \gamma_\nu\}=2\delta_{\mu\nu}$~(see, for instance, Refs.~\cite{Shimizu,Chuy,Edward}). In any of these representations, it is impossible to define chiral symmetry~\cite{Appel,Appelquist,Pisarski,Semenoff}. Furthermore, the fermion mass term, regardless of its origin, is parity non-invariant. Still, one can construct a parity-preserving Lagrangian considering
two different species with a relative sign between their masses~\cite{Shimizu,Chuy}. As a result, two independent chiral transformations can be defined~\cite{Chuy}. These two species are conveniently merged into a single four-component spinor, making use of a reducible representation of the Dirac matrices.   As compared with QED in (3+1)-dimensions, only three Dirac
matrices are required to describe the dynamics of planar fermions. Once we have selected a set of matrices, say $\{\gamma_0,\gamma_1,\gamma_2\}$, to write down the Dirac equation, two anti-commuting gamma matrices, namely, $\gamma_3$ and $\gamma_5$ remain unused. Hence the corresponding massless Dirac Lagrangian is invariant under the chiral-like transformations $\psi\to e^{i\alpha\gamma_3}\psi$ and $\psi\to e^{i\beta\gamma_5}\psi$,   that is, it is invariant under a global $U(2)$ symmetry with generators $1$, $\gamma^3$, $\gamma^5$ and $[\gamma^3, \gamma^5]$, corresponding to the interchange of
fermion species.  This symmetry is explicitly broken by an ordinary mass term $m_e\bar\psi\psi$. The order parameter of spontaneous symmetry
breaking is the condensate $\langle \bar\psi\psi \rangle_e=\langle 0|\bar\psi\psi |0\rangle.$ {Although this definition suggests that the condensate is a constant mass-scale which fills all space-time, the modern perspective suggests that condensates are, in fact, a property of bound states generated by the theory under study~\cite{brodsky}. Note that there exists a second mass term -- $m_o \bar\psi \tau \psi$, with $\tau=[\gamma_3,\gamma_5]/2=diag(I,-I)$ -- which is invariant under the ``chiral'' transformations. This term is sometimes referred to as the Haldane mass term~\cite{Haldane} and is associated with the condensate $\langle\bar\psi\psi\rangle_o= \langle0|\bar\psi\tau\psi|0\rangle$. Moreover, the ordinary mass term  is even under parity transformations, but the Haldane mass term is not. This justifies the use of subscripts $e$ for even and $o$ for odd in the respective quantities of the model.

It is known that the Haldane mass term radiatively induces a parity-odd  contribution into the vacuum polarization~\cite{CH,Khare}, which can be traced back to an induced Chern-Simons interaction of the form
\be
{\cal L}_{CS}=-\frac{i\theta}{4}\varepsilon_{\mu\nu\rho}A_\mu F_{\nu\rho}\;.\label{csl}
\ee
Such a term is parity non-invariant, and despite the fact that it is not manifestly gauge invariant, the corresponding action respects gauge symmetry. The parameter $\theta$ induces a topological mass for the photons. The model we consider in this paper is  described by the Lagrangian 
\bea
{\cal L}&=& \bar\psi (i\slsh{\partial}+e\slsh{A}+m_e+\tau m_o)\psi
+\frac{1}{4}F_{\mu\nu}F_{\mu\nu}\nn\\&&
+\frac{1}{2\xi}(\partial_\mu A_\mu)^2-\frac{i\theta}{4}\varepsilon_{\mu\nu\rho}A_\mu F_{\nu\rho}.\label{mcslag}
\eea
Written in this form, neither $m_e$ nor $m_o$ correspond to poles of the fermion propagator, and hence these masses cannot be associated with a specific fermion species. Nevertheless, introducing the chiral projectors $\chi_\pm = (1\pm \tau)/2 $, which have the properties  $\chi_\pm^2=\chi_\pm\;, \chi_+\chi_-=0\;, \chi_++\chi_-=1$, we may define right-handed $\psi_+$ and left-handed $\psi_-$ fermion fields as $\psi_\pm=\chi_\pm \psi$. The $\chi_\pm$ project the upper and lower two-component spinors (fermion species) out of the four-component $\psi$, in such a manner that the fermion sector of the Lagrangian~(\ref{mcslag}) can be cast into the form
\be
{\cal L}_F= \bar\psi_+ (i\slsh{\partial}-m_+)\psi_++\bar\psi_- (i\slsh{\partial}-m_-)\psi_-,\label{fermchi}
\ee
with $m_\pm=m_e\pm m_o$. This Lagrangian explicitly describes two fermion species of physical masses $m_+$ and $m_-$, respectively. These masses break chiral symmetry and parity at the same time. Moreover, the parity-violating mass removes the mass degeneracy between the two species. In what follows, we are interested in the analytical properties of the dynamically generated fermion propagator associated with the physical masses.

\section{Gap equation}
\label{DynamicalFM}

The analytical structure of the propagator can be studied with the corresponding Schwinger-Dyson equation,
\bea
S_F(p)^{-1}&=&S_F^{(0)}(p)^{-1}\nn\\&&\hspace{-8mm}
+\ e^2 \int \frac{d^3k}{(2\pi)^3} \Gamma^\mu(k, p) S_F(k) \gamma^\nu \Delta_{\mu\nu}(k-p)\;,\label{SDE}
\eea
where $\Gamma^\mu(k,p)$ and $\Delta_{\mu\nu}(k-p)$ are, respectively, the full fermion-photon vertex and the full photon propagator, which themselves obey their own SDE. The quantities $S_F(p)$ and $S_F^{(0)}(p)$ stand for the full and bare fermion propagator. In QED$_3$, the coupling $e^2$ has mass-dimension one. Moreover, since the theory is super-renormalizable, $e^2$ becomes the natural scale of massless QED$_3$, which is directly connected to the scales of confinement and dynamical chiral symmetry breaking. In this article, we write all mass scales in units of $e^2=1$.

From the Lagrangian~(\ref{mcslag}), we deduce that the inverse fermion propagator has the form
\be
S_F^{-1}(p)=  A_e(p)\slsh{p} +A_o(p) \tau \slsh{p}-B_e(p)-B_o(p)\tau\;.
\ee
The scalar functions $A_{e,o}(p)$ and $B_{e,o}(p)$ can be expressed in terms of the fermion wavefunction renormalizations $F_{e,o}(p)$ and the mass functions $M_{e,o}(p)$ as $A_{e,o}(p)=1/F_{e,o}(p)$ and $B_{e,o}(p)=M_{e,o}(p)/F_{e,o}(p)$, both in the even and odd sectors. The bare propagator corresponds to the values $A_e^{(0)}(p)=1,\,A_o^{(0)}(p)=0,\,B_e^{(0)}(p)=m_e,\,B_o^{(0)}(p)=m_o$. Rather than working with parity eigenstates, we find it
convenient to work with the chiral  Lagrangian~(\ref{fermchi}). The chiral decomposition of the fermion propagator then becomes
\bea
S_F(p)&=& -\frac{A_+ (p)\slsh{p}+B_+(p)}{A_+^2(p)p^2+
B_+^2(p)}\chi_+ \nn\\&&
- \frac{A_- (p)\slsh{p}+B_-(p)}{A_-^2(p)p^2+B_-^2(p)}\chi_- 
\nn\\
&\equiv& -[\sigma_+^V(p^2) \slsh{p}+\sigma_+^S(p^2)] \chi_+\nn\\
&&-[\sigma_-^V(p^2) \slsh{p}+\sigma_-^S(p^2)] \chi_-
\;,\label{fpp}
\eea
where our notation is  $K_\pm=K_e\pm K_o$ for all relevant even and odd quantities of the model. The inverse transformations are simply $K_e=(K_++K_-)/2$ and $K_o=(K_+-K_-)/2$. In this basis, the bare quantities are $A_\pm^{(0)}(p)=1$ and $B_\pm^{(0)}(p)=m_\pm$. 

We now consider vacuum polarization effects. Assuming $N$ different fermion families, initially massless, we consider the one-loop vacuum polarization of the photon at leading order in the $1/N$ expansion. The vertex remains bare in this truncation. In the absence of a CS term, it has been found that there exists a critical value $N_c=32/\pi^2$ above which chiral symmetry is restored~\cite{Appelquist}. The reliability of the truncation, however, has been challenged in Ref.~\cite{Pennington} by the possibility of an exponentially suppressed dynamical mass for arbitrarily large values of $N$. More recently, it was established that QED$_3$ can have a critical number of fermion families for chiral symmetry breaking {\em iff} the wave-function renormalization and vacuum polarization are homogeneous functions at infrared momenta when the fermion mass function
vanishes~\cite{BCRR}. Whether such a critical value $N_c$ exists, depends on the exact form of the fermion-photon vertex, which simply relates the anomalous dimensions of the wave-function renormalization and vacuum polarization through the Ward identity. In this connection, lattice-QED$_3$ suggests that $N_c>1$~\cite{hands}, with the recently reported value of $N_c\sim 1.5$~\cite{kogut}, in agreement with the upper bound $N_c\le 3/2$, derived by thermodynamic arguments~\cite{AppTL}. Including the CS term, chiral symmetry restoration within this truncation scheme has been explored in great detail in Refs.~\cite{KondMarPRL,KondMarPRD}. Here we want to explore the consequences of this phase transition for confinement.
Defining the new coupling $\tilde{\alpha}=N/8$, at leading order of the $1/N$ expansion including the CS term, the photon propagator  acquires the form
\begin{eqnarray}
 \Delta_{\mu\nu}(q)&=&\frac{q^{2}+\tilde{\alpha}|q|}{q^{2}[(|q|+\tilde{\alpha})^{2}+\theta^{2}]}\Big(\delta_{\mu\nu}
-\frac{q_{\mu}q_{\nu}}{q^{2}}\Big)\nn\\
&-&\frac{\epsilon_{\mu\nu\rho}q_{\rho}\theta}{q^{2}[(|q|+\tilde{\alpha})^{2}+\theta^{2}]},
\end{eqnarray}
where the Landau gauge $\xi=0$ was chosen.

Inserting this propagator into the SDE for the fermion propagator, Eq.~(\ref{SDE}), in the chiral basis, we obtain the following system of equations
\begin{eqnarray}
 A_{\pm}(p) &=& 1\nn\\
&+& \frac{16\tilde{\alpha}}{N p^{2}}\int  \frac{d^{3} k}{(2\pi)^{3}}\sigma_\pm^V(k^2) 
\frac{(k\cdot q)(p\cdot q) (q^{2}+\tilde{\alpha} |q|)}{q^{4}[(|q|+\tilde{\alpha})^{2}+\theta^{2}]} \nonumber \\
 &\mp& \frac{16\tilde{\alpha} \theta}{ N p^{2}}\int  \frac{d^{3} k}{(2\pi)^{3}}\sigma_\pm^S(k^2) 
\frac{(p\cdot q)}{q^{2}[(|q|+\tilde{\alpha})^{2}+\theta^{2}]}\nn \\
 B_{\pm}(p) &=&  \frac{16\tilde{\alpha}}{ N}\int  \frac{d^{3} k}{(2\pi)^{3}}\sigma_\pm^S(k^2) 
\frac{q^{2}+\tilde{\alpha}|q|}{q^{2}[(|q|+\tilde{\alpha})^{2}+\theta^{2}]} \nonumber \\
 &\mp& \frac{16\tilde{\alpha} \theta}{ N}\int  \frac{d^{3} k}{(2\pi)^{3}}\sigma_\pm^V(k^2) 
\frac{(k\cdot q)}{q^{2}[(|q|+\tilde{\alpha})^{2}+\theta^{2}]},\label{sde1n}
\end{eqnarray}
with $q=k-p$. As expected, $A_\pm(p)\simeq 1+{\cal O}(1/N)$, and hence the influence of wave-function renormalization on $B_\pm(p)$ is subdominant. We now solve the system of Eqs.~(\ref{sde1n}) by varying the parameters $N$ and $\theta$. 
\begin{figure}
\begin{center}
\includegraphics[width=0.35\textwidth, angle=270]{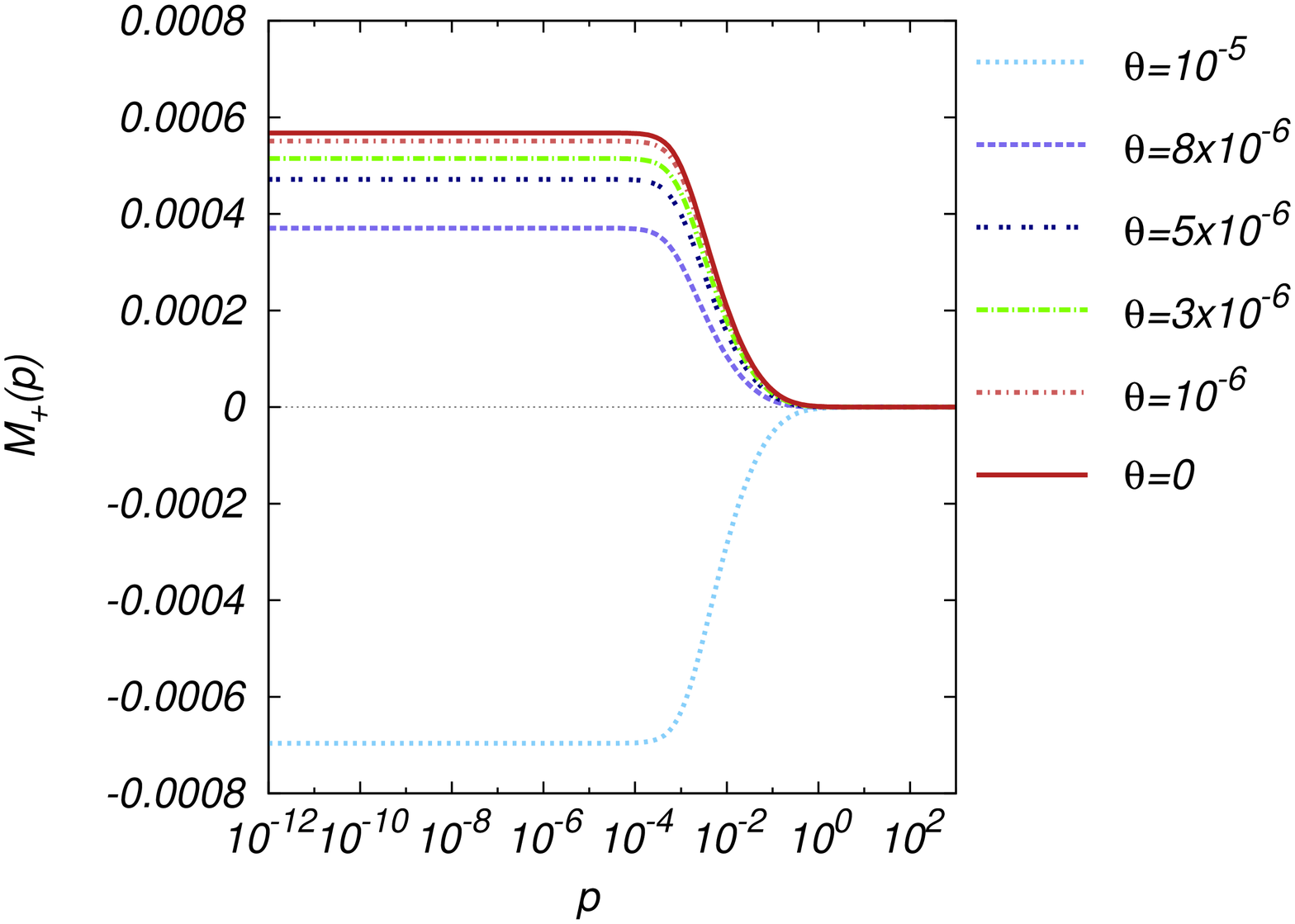}
\includegraphics[width=0.35\textwidth, angle=270]{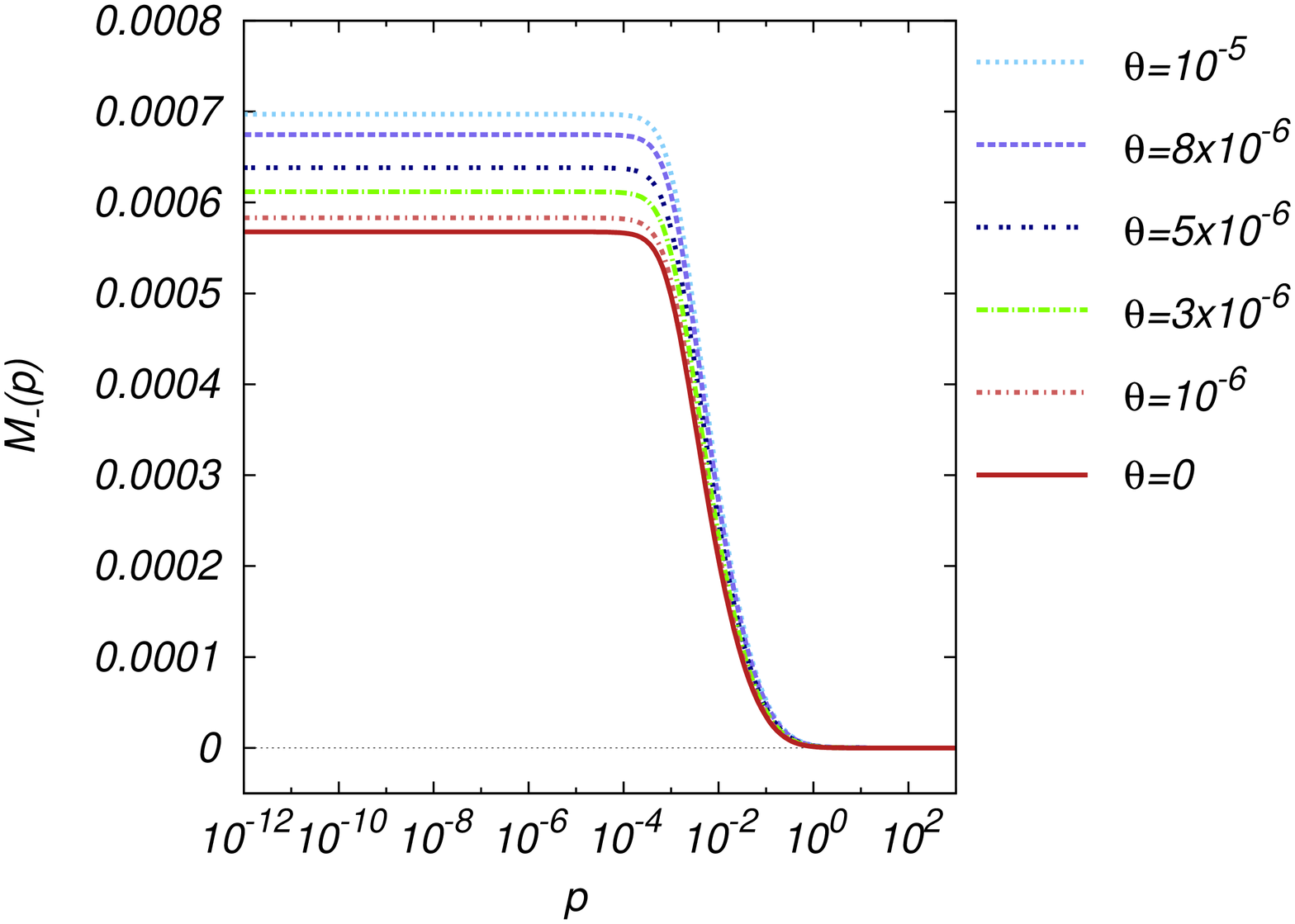}
\caption{The mass functions $M_\pm(p)$ for $N=2$ and for various values of the Chern-Simons parameter $\theta$. Departing from the parity-preserving value, $\theta=0$, $M_+(p)$ diminishes in height whereas $M_-(p)$ increases with $\theta$. At $\theta_c\sim 8\times 10^{-6}$, the plateau of $M_+(p)$ turns to negative values, whereas $M_-(p)$ continues increasing.}
\label{fig1nmpm}
\end{center}
\end{figure}

For fixed $N$ ($N<N_c$), the solutions exhibit a discontinuity in the behavior of the propagator at some critical value $\theta_c$ which depends on $N$. In Fig.~\ref{fig1nmpm}, we show the mass functions $M_\pm(p)$ for various values of $\theta$ and $N=2$. We observe that the height of $M_-(p)$ increases from the parity-preserving value as $\theta$ increases, whereas $M_+(p)$ decreases. At $\theta_c\simeq 8\times 10^{-6}$, a discontinuity takes place and the height of $M_+(p)$ becomes negative. This sudden drop of the function $M_+(p)$ has strong implications regarding chiral symmetry restoration of the model~\cite{KondMarPRL,KondMarPRD}. In the infrared, the height of the plateau can be considered as an order parameter for dynamical chiral symmetry breaking. In this connection, in Fig.~\ref{fig1nmu}, we draw the dependence of $\mu_\pm=M_\pm(0)$ on the parameter $\theta$ below and above criticality. The quantity $\mu_\pm$ can be regarded as the dynamical mass of the corresponding fermion species. We observe that the role of the CS coefficient is to remove the mass degeneracy between the two fermion species as long as $\theta<\theta_c$ -- there is a light and a heavy species. At $\theta_c$, however, there is a drastic change in this behavior: the light species develops a negative mass, due to the fact that each species behaves effectively as an irreducible fermion, for which the mass term is a pseudoscalar under parity transformations and hence the corresponding mass coefficient is not restricted to take only positive values. Such a mass has the same magnitude as the mass of its heavy cousin, as can be appreciated in the dotted curve in the same graph. This implies that above $\theta_c$, the would-be parity preserving mass $\mu_e$, which is obtained by inverting the chiral transformation as $\mu_e=(\mu_++\mu_-)/2=0$, and hence chiral symmetry is restored~\cite{KondMarPRL,KondMarPRD}. Nevertheless, we want to emphasize that in this model, $\mu_e$ does not correspond to a pole in a propagator and hence to a physical mass. Physical masses $\mu_\pm$ are generated for arbitrarily large values of $\theta$.
\begin{figure}
\begin{center}
\includegraphics[width=0.35\textwidth, angle=270]{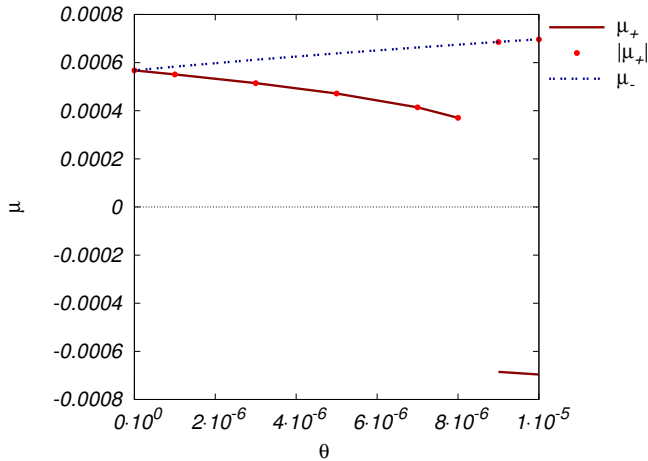}
\caption{$\mu_\pm$ as a function of $\theta$ for $N=2$. Both masses are equal at $\theta=0$ and start deviating as $\theta$ increases. $\mu_-$ becomes heavier and $\mu_+$ lighter. At $\theta_c$, $\mu_+$ becomes negative, with the same magnitude as ${\mu_-}$.}
\label{fig1nmu}
\end{center}
\end{figure}
\begin{figure}
\begin{center}
\includegraphics[width=0.35\textwidth, angle=270]{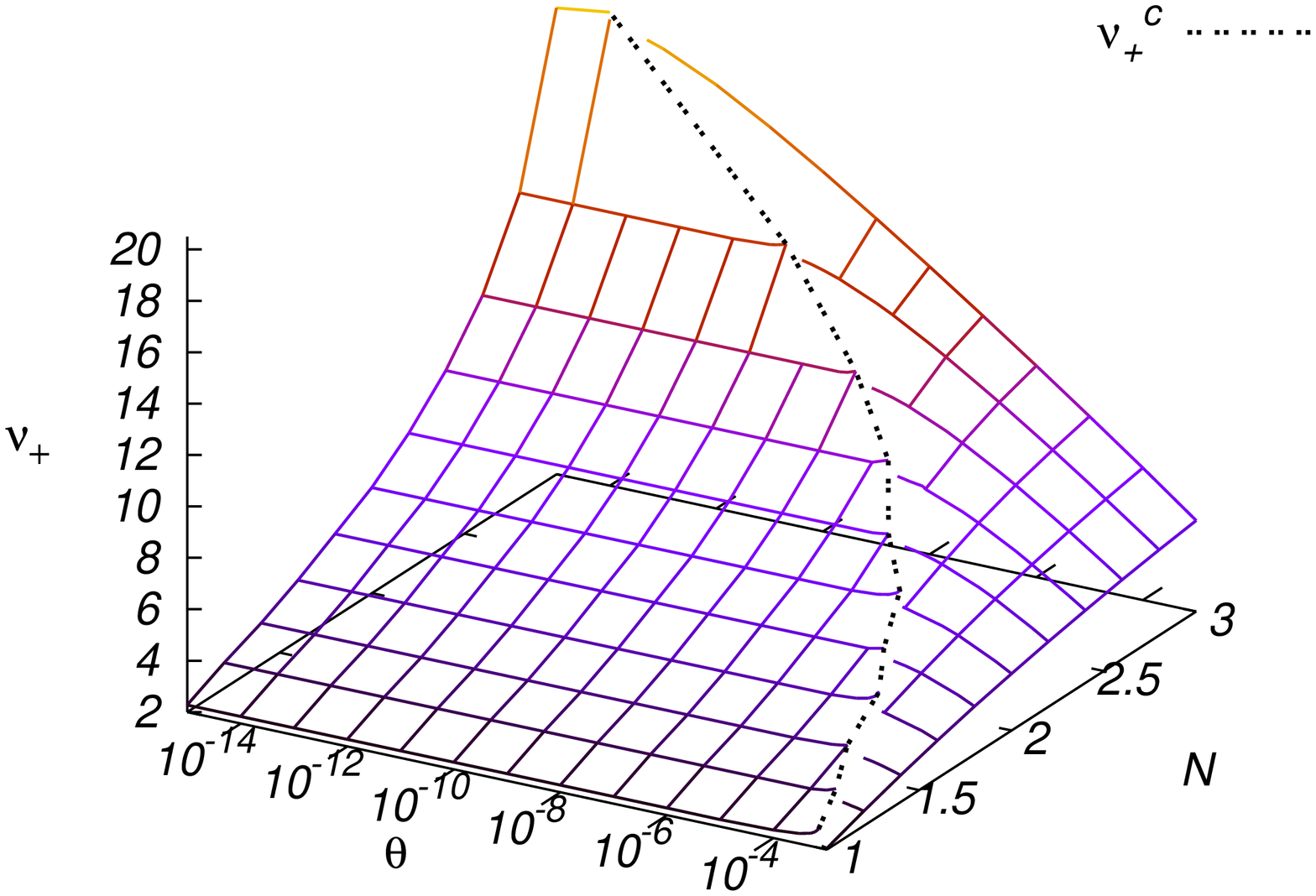}
\includegraphics[width=0.35\textwidth, angle=270]{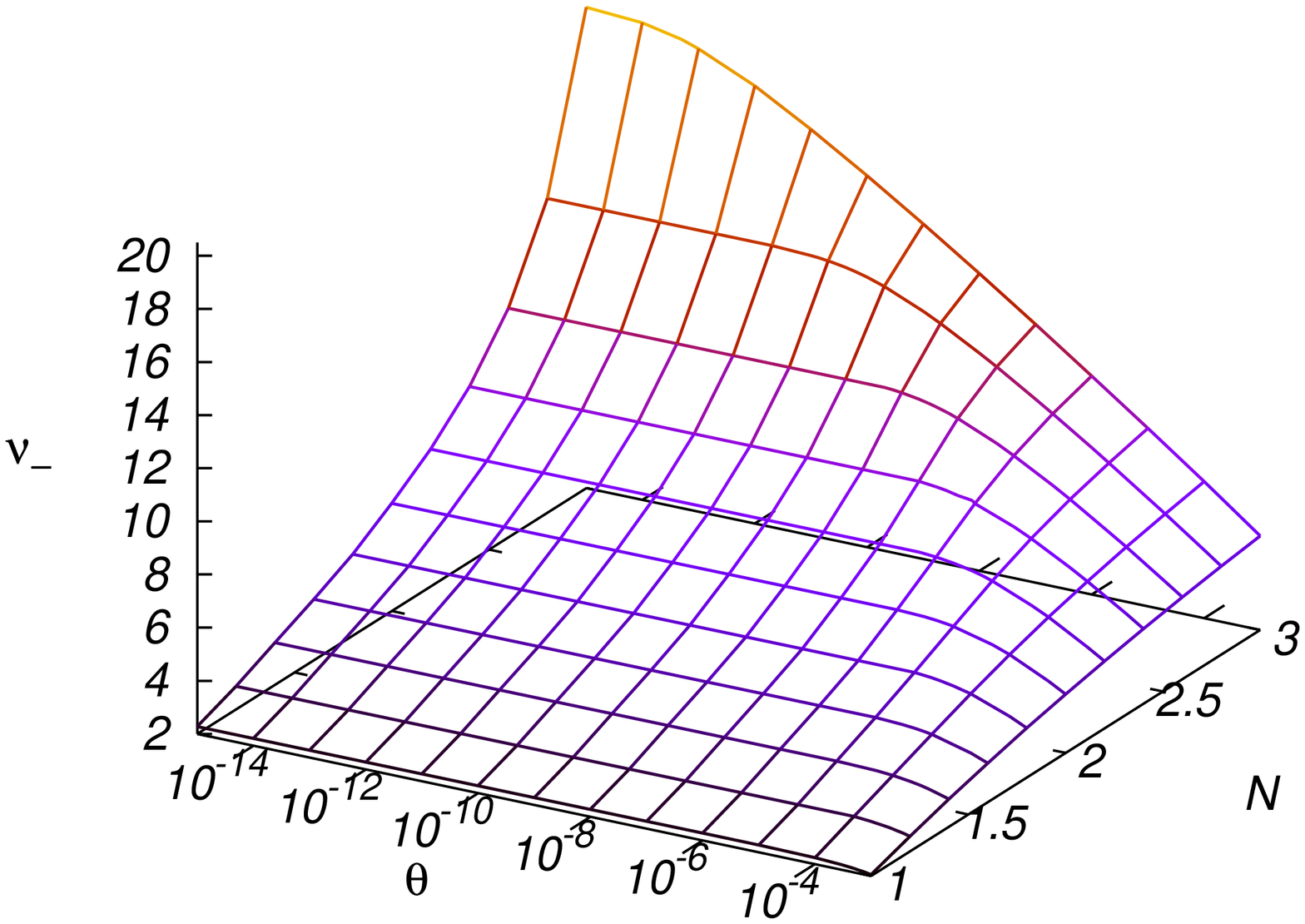}
\caption{$\nu_+$  and $\nu_-$, defined in Eq.~(\ref{nus}), as a function of $N$ and $\theta$. {\em Upper panel:} $\nu_+$ increases for larger $N$. Along the curve $\nu_+^c$ which specifies the critical value of $\theta$ for a given $N$, it suffers a discontinuity. {\em Lower panel:} $\nu_-$ evolves smoothly. It diverges in the region where $N\to N_c$ for small values of $\theta$.}
\label{fig1n3d}
\end{center}
\end{figure}

Let us define the parameters
\be
\nu_\pm =-\ln \left| \frac{\mu_\pm}{\tilde{\alpha}}\right|\;.\label{nus}
\ee
The dependence of  $\nu_+$  (upper panel) and $\nu_-$ (lower panel) on the parameters $N$ and $\theta$ is shown in Fig.~\ref{fig1n3d}. For $\nu_+$, the dotted curve describes the critical curve $\nu_+^c$, specified by the pair of values $(N_c,\theta_c)$, where a first-order phase transition takes place. Here it is evident that $N_c$ depends on $\theta$. Moreover, the data are consistent with the numerical fit advocated in Ref.~\cite{KondMarPRD},
\be
\theta_c\simeq \exp{\left[\frac{-A+\delta}{\sqrt{\frac{N_c(0)}{N}-1}}\right]}\;,
\ee
where $e^\delta$ represents an amplitude and $A$ a damping factor, both functions of $N$.
On the other hand, $\nu_-$ varies smoothly with $N$ and $\theta$. Notice, that  masses vanish at the point $(N=N_c(0),\theta=0)$, where $\nu_\pm\to \infty$. In this regime, the masses $\mu_+=\mu_-=\mu$ follow the well-known behavior~\cite{Appel,Appelquist} of ordinary QED$_3$ in this approximation, 
\be
\mu \simeq  \tilde\alpha \exp{\left[\frac{-2\pi+\delta'}{\sqrt{\frac{N_c(0)}{N}-1}} \right]}.
\ee
In the $(N,\theta)$-plane, fermions remain massless along the segment $(N>N_c(0),\theta=0)$. For any finite value of $\theta$, but arbitrary $N$, both $\mu_\pm$ are finite too, and thus the fermions are massive. This is illustrated in Fig.~\ref{figN4}, where $\mu_-$ and $|\mu_+|$ for $N=4$ are shown as a function of $\theta$. Since the magnitude of these masses is the same, there is no mass associated with chiral symmetry breaking $\mu_e$. Only the Haldane mass term contributes to $\mu_\pm$.

\begin{figure}
\begin{center}
\includegraphics[width=0.35\textwidth, angle=270]{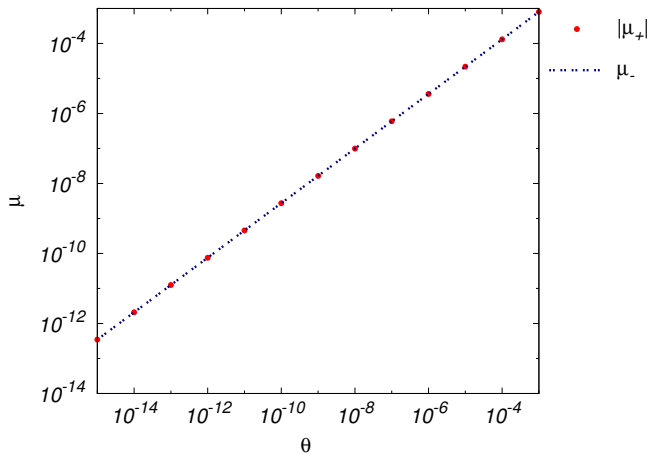}
\caption{$\mu_\pm$ as a function of $\theta$ for $N=4$. For this value of $N$, the only contribution to the masses comes from the Haldane mass. The masses $\mu_+$ and $\mu_-$ have the same magnitude, but are opposite in sign.}
\label{figN4}
\end{center}
\end{figure}

In the next section, we shall consider the implications of this phase transition for confinement. 

\section{Confinement}
\label{confinement}

Quenched compact QED$_3$ without a Chern-Simons term possesses nonzero string tension~\cite{conf}, and thus is confining. This feature persists in the unquenched theory if fermions of either explicit or dynamical mass circulate in the photon vacuum polarization~\cite{Burden}, because only massless fermions can completely screen charges. For general implications of confinement the reader may consult Refs.~\cite{Craigconf1,Craigconf2,Craigconf3,Craigconf4,Confosc1,Confosc2}. Specifically for QED$_3$, the subject has been discussed, e.g., in Refs.~\cite{Maris96,BRfbs}. In particular, conditions for the chiral symmetry restoration and confinement/deconfinement phase transition to occur simultaneously are presented in Refs.~\cite{BCRR,BRRS}. Here we are interested in the extent up to which confinement persists in the presence of a CS term.

Whether a solution of the SDE supports confinement can be tested through the violation of the Osterwalder-Schrader axiom of reflection positivity~\cite{ost1,ost2}, which states that the spatially averaged Schwinger function,
\be
\Delta(t)=\int d^2 x \int \frac{d^3 p}{(2\pi)^3}e^{i(t p_{o}+x\cdot p)}\sigma_s(p^2)\;,\label{spati}
\ee
should be positive definite if it is related to a stable asymptotic state~\cite{Confosc1,Confosc2}. In our case, we construct the functions $\Delta_\pm(t)$ by inserting the solutions of the SDE into the above expression.  In Fig.~\ref{figConfOsc} we plot these functions,
\begin{figure}
\begin{center}
\includegraphics[width=0.35\textwidth, angle=270]{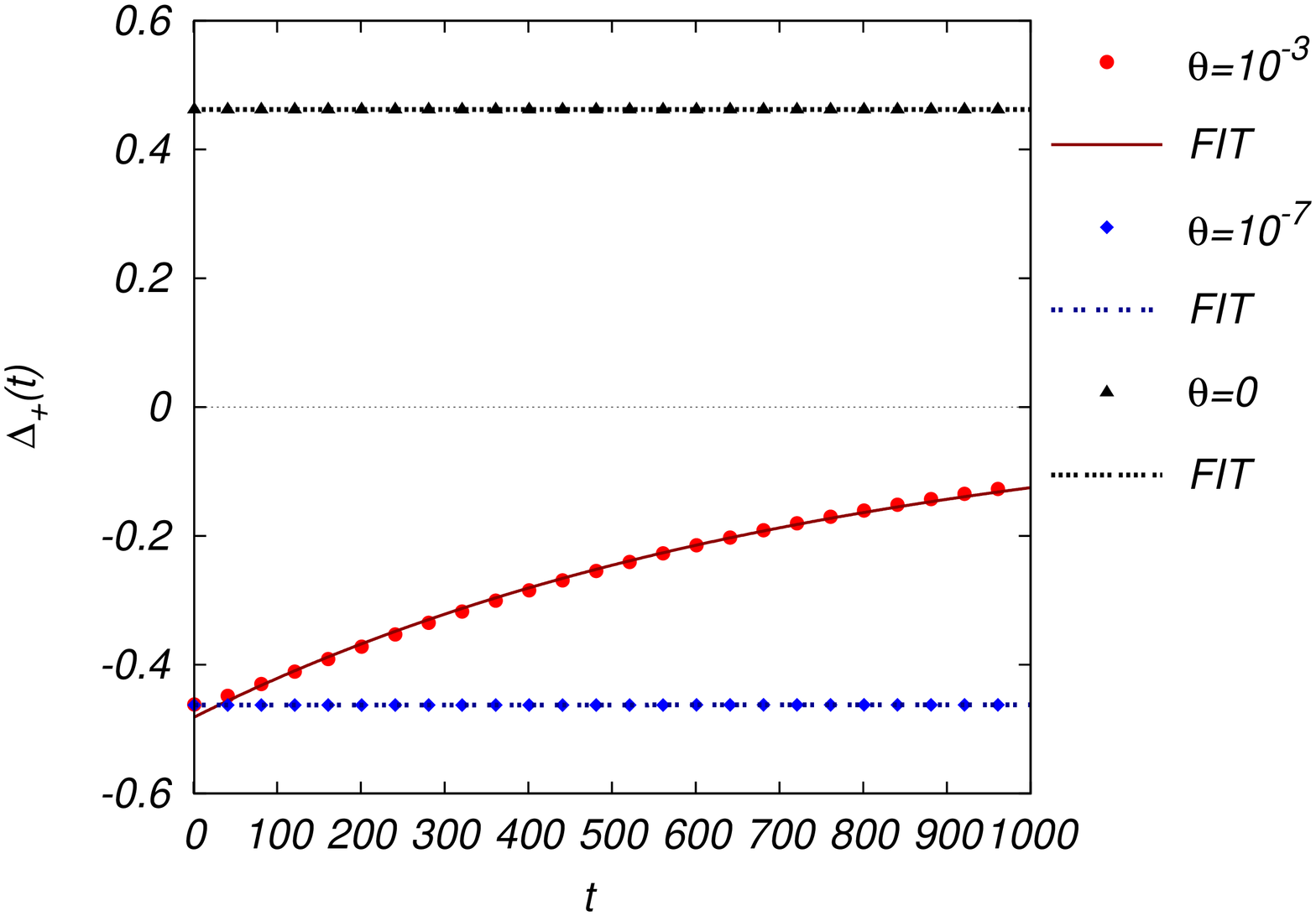}
\includegraphics[width=0.35\textwidth, angle=270]{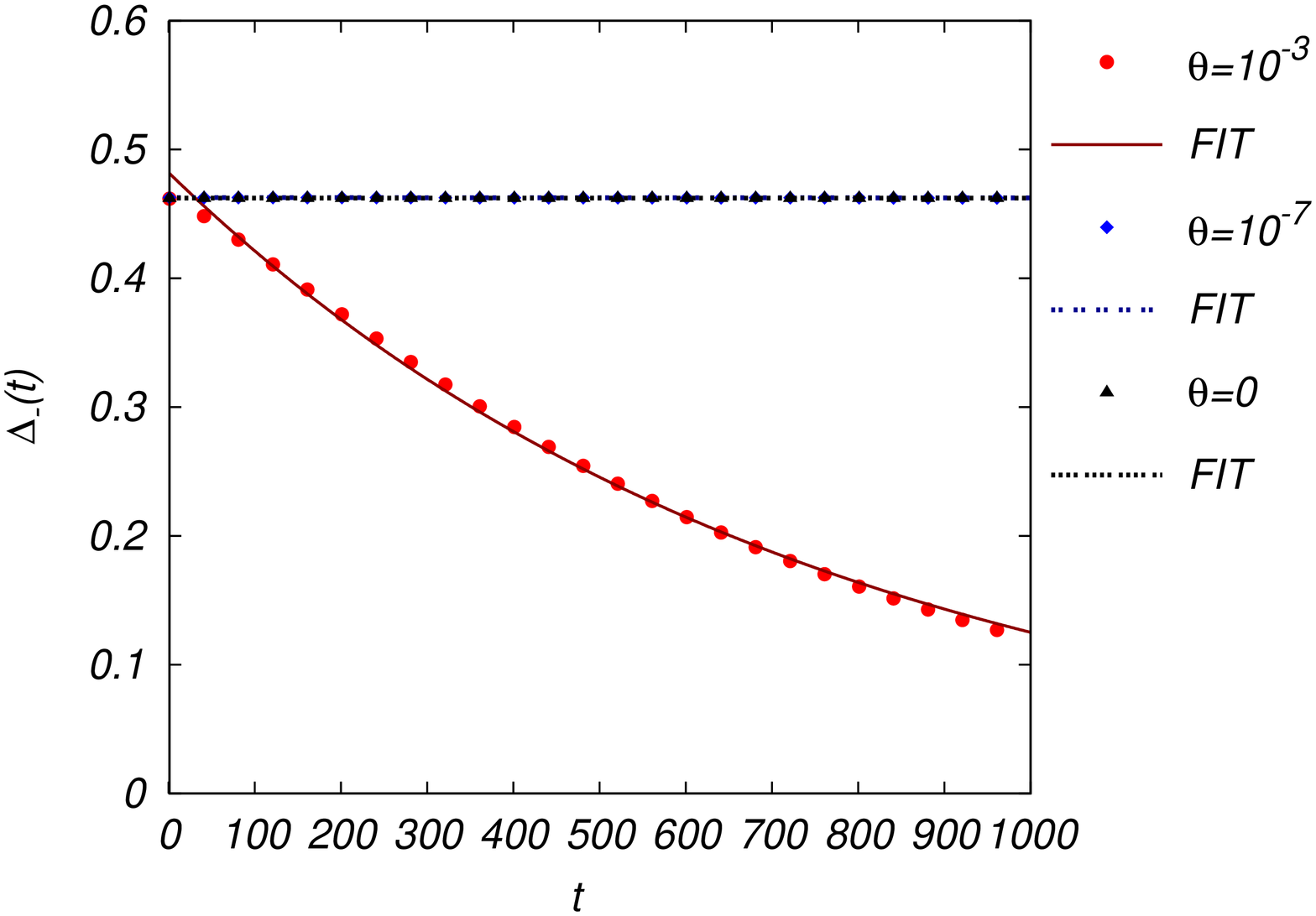}
\caption{Spatially averaged Schwinger functions for $N=3$ and various values of $\theta$  fitted according to Eq.~(\ref{fitfree}). {\em Upper panel:} Below $\theta_c$, $\Delta_+(t)$ is positive definite, and becomes negative definite above criticality. {\em Lower panel:} $\Delta_-(t)$ is always positive definite. These functions describe deconfined fermions.}
\label{figConfOsc}
\end{center}
\end{figure}
using a numerical fit of the form 
\be
\Delta_\pm(t)=\frac{1}{2}{\rm sgn}(\mu_\pm) e^{-|\mu_\pm| t}.\label{fitfree}
\ee 
Because below $\theta_c$, the quantities $\mu_\pm$ are positive, these Schwinger functions describe stable excitations of masses $\mu_\pm$, respectively, i.e, fermions in this truncation are deconfined (see for instance, Ref.~\cite{Maris96}).  Deconfinement with dynamical masses is a feature of the leading $1/N$ truncation, since fermions in the vacuum polarization are effectively massless. For $\theta\ge\theta_c$, on the other hand, $\mu_->0$ but  $\mu_+=-\mu_-$ and thus $\Delta_+(t)$ is no longer positive definite, but becomes negative definite. However, because this function does not change sign as $t\to \infty$, it does not violate the axiom of reflexion positivity, and thus fermions remain deconfined.
\begin{figure}
\begin{center}
\includegraphics[width=0.35\textwidth, angle=270]{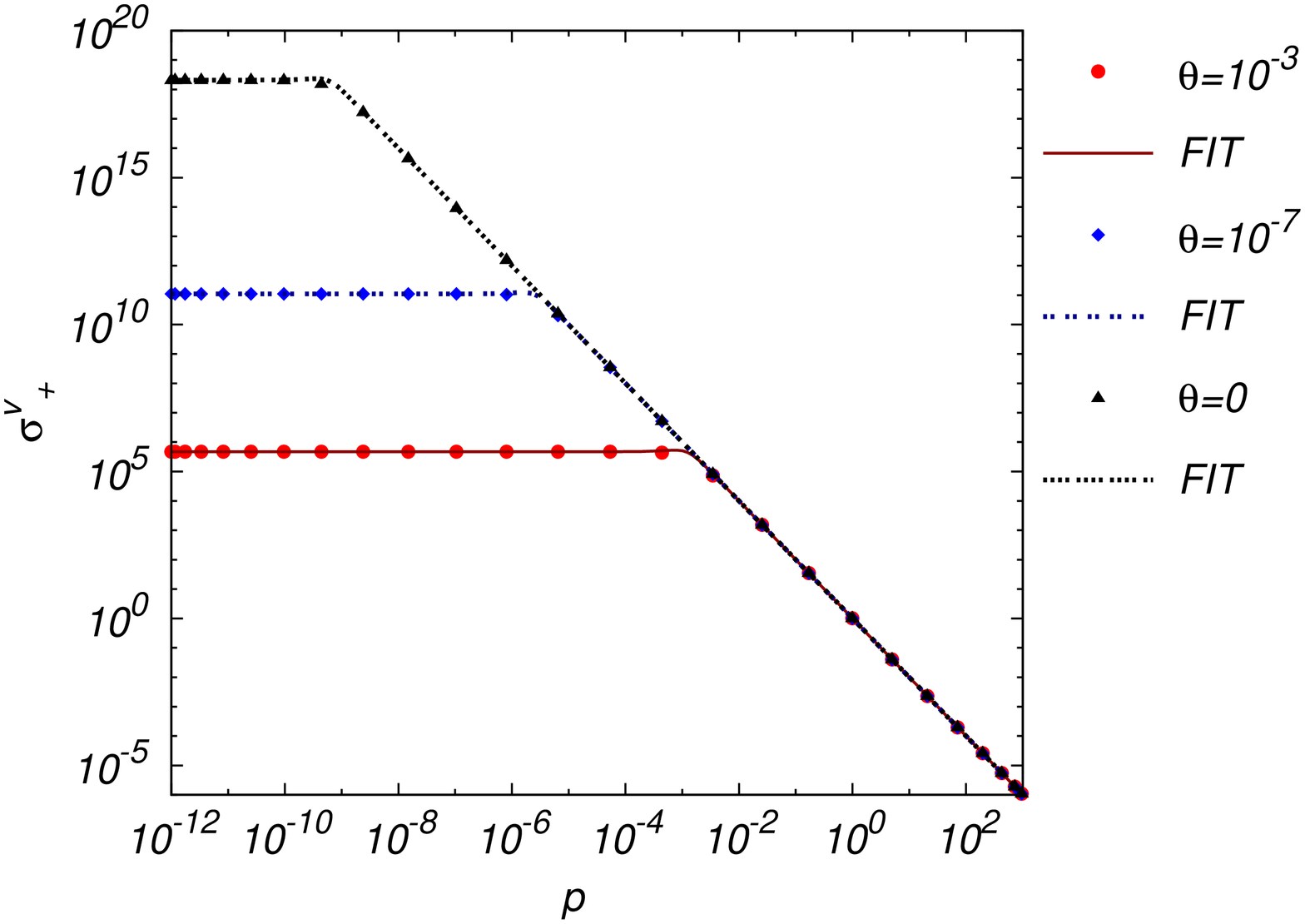}
\includegraphics[width=0.35\textwidth, angle=270]{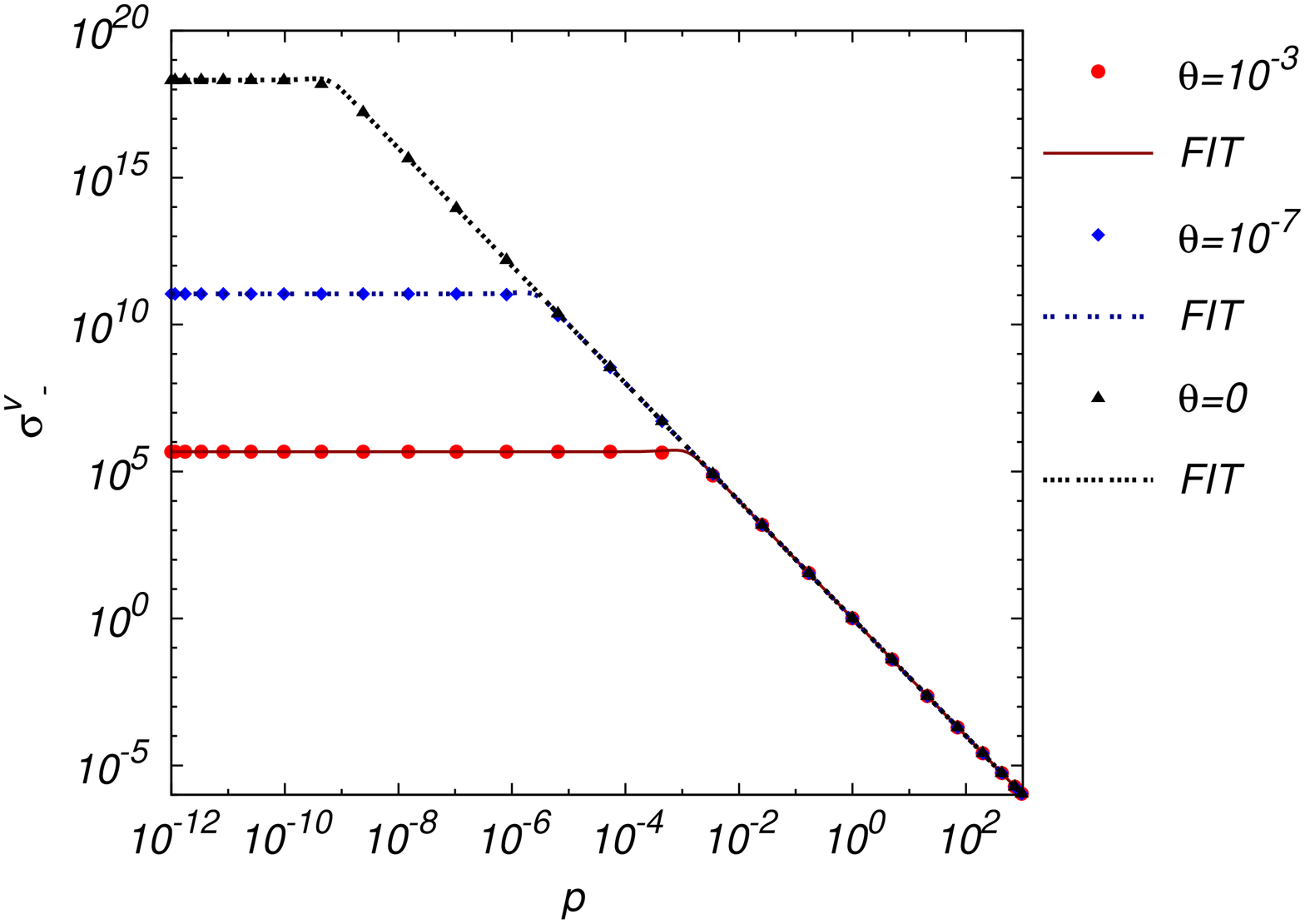}
\caption{Comparison of the vector parts of the fermion propagator in the chiral basis for various values of $\theta$ at fixed $N=3$ with the fit in Eq.~(\ref{fit}).}
\label{fig1n4}
\end{center}
\end{figure}

\begin{figure}
\begin{center}
\includegraphics[width=0.35\textwidth, angle=270]{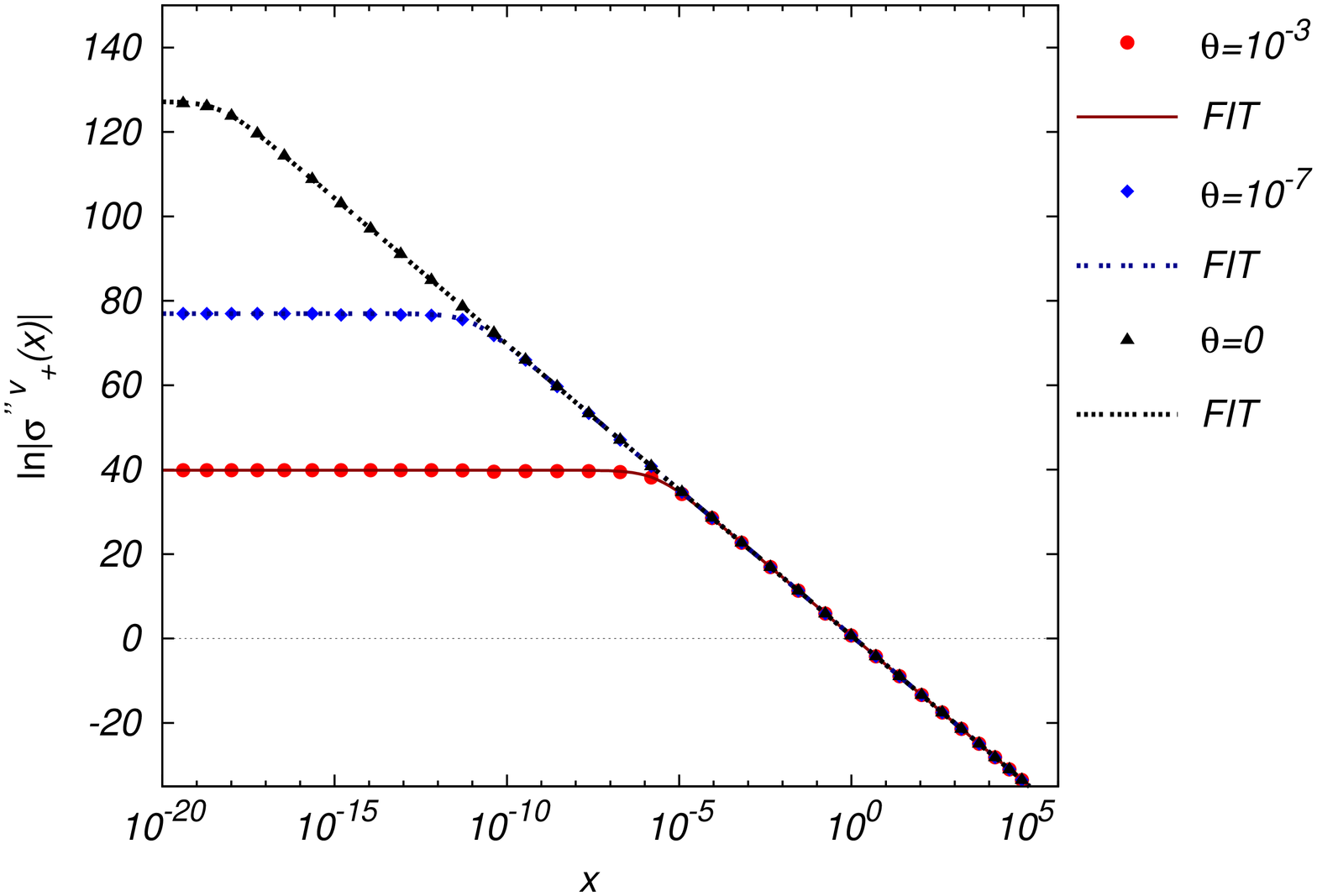}
\includegraphics[width=0.35\textwidth, angle=270]{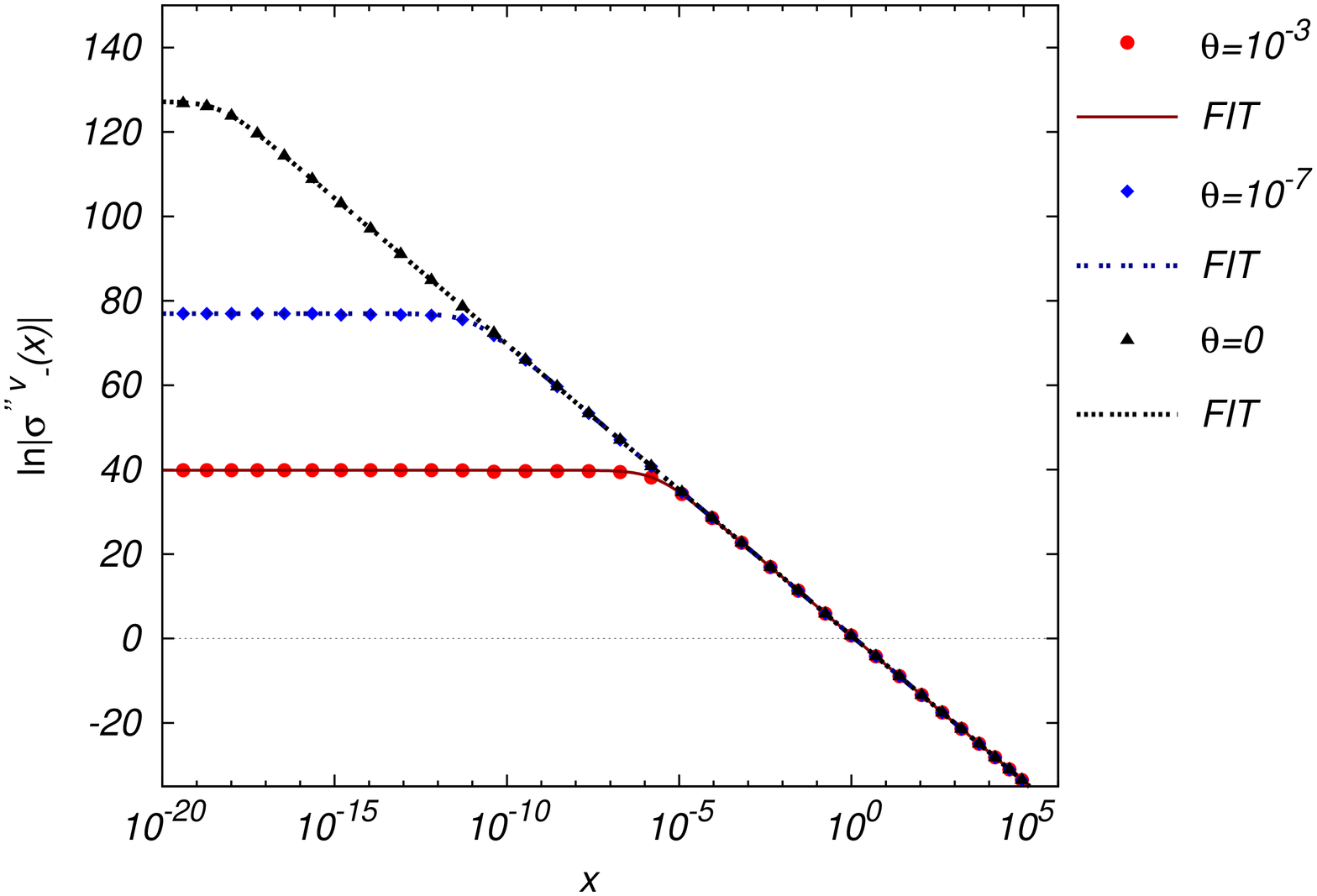}
\caption{Comparison of the second derivatives of the vector parts of the fermion propagator as a function of $x=p^2$ for $N=3$ and $\theta=10^{-3}$ with the fit in Eq.~(\ref{fit}). }
\label{figinflex}
\end{center}
\end{figure}

The above confinement test can only be performed in the massive phase of the theory, where $\sigma_s(p^2)\ne 0$. In some situations, however, it is desirable to have the means to explore the scenario of confinement without resorting to dynamical mass generation properties of the propagator. This can be achieved by recalling that any Schwinger function with an inflexion point at some $p^2>0$ must violate the axiom of reflexion positivity. In Fig.~\ref{fig1n4} we plot the functions $\sigma_\pm^V(p^2)$  for $N=3$ and different values of $\theta$, above and below chiral symmetry restoration. These show a behavior that is typical of the vector part of a massive fermion propagator.
Following Refs.~\cite{BCRR,BRRS}, we set $x=p^2$. Thus, the order parameters for confinement are defined as the inflection points $x^c_\pm$ of $\sigma_\pm^V(x)$, namely, points at which
\be
\left.\frac{d^2}{dx^2}\sigma_\pm^V(x)\right|_{x=x_\pm^c}=0.
\ee
From our numerical results, we observe that in the present case, the vector parts of the fermion propagator do not develop inflection points, as shown in Fig.~\ref{figinflex}, where the logarithm of the second derivatives of $\sigma_\pm^V(x)$ for $N=3$ are plotted as a function of $x$ for various values of $\theta$, above and below $\theta_c$.  This statement can be further verified by noticing that these
functions can be fitted as
\be
\sigma_\pm^V(x)=\frac{1}{x+\mu_\pm^2},\label{fit}
\ee
for arbitrary values of $\theta$. Thus, a simple exercise reveals that the second derivative of the above fit never changes sign, and thus confinement is absent in this propagator. This is also shown in Fig.~\ref{figinflex}.

Alternatively, one can look at the global properties of $\sigma_\pm^V(x)$ and carry out a similar analysis of that in Eq.~(\ref{spati}). We define  new spatially averaged Schwinger functions, now involving the vector parts of the propagator, namely,
\be
\Omega_\pm(t)=\int d^2 x \int \frac{d^3 p}{(2\pi)^3}e^{i(t p_{o}+x\cdot p)}\sigma_\pm^V(x)\;.\label{spativec}
\ee
In Fig.~\ref{testsigmav} we present the logarithm of the absolute values of $\Omega_\pm(t)$ for $N=3$ and various values of $\theta$. We further make a comparison with the fit
\be
\Omega_\pm(t)= \frac{1}{2|\mu_\pm|} e^{-|\mu_\pm|t}\;.\label{fitomega}
\ee
We observe that the functions  $\Omega_\pm(t)$ are always positive definite, even after chiral symmetry is restored. Therefore, these describe fermions in a deconfined phase.
\begin{figure}
\begin{center}
\includegraphics[width=0.35\textwidth, angle=270]{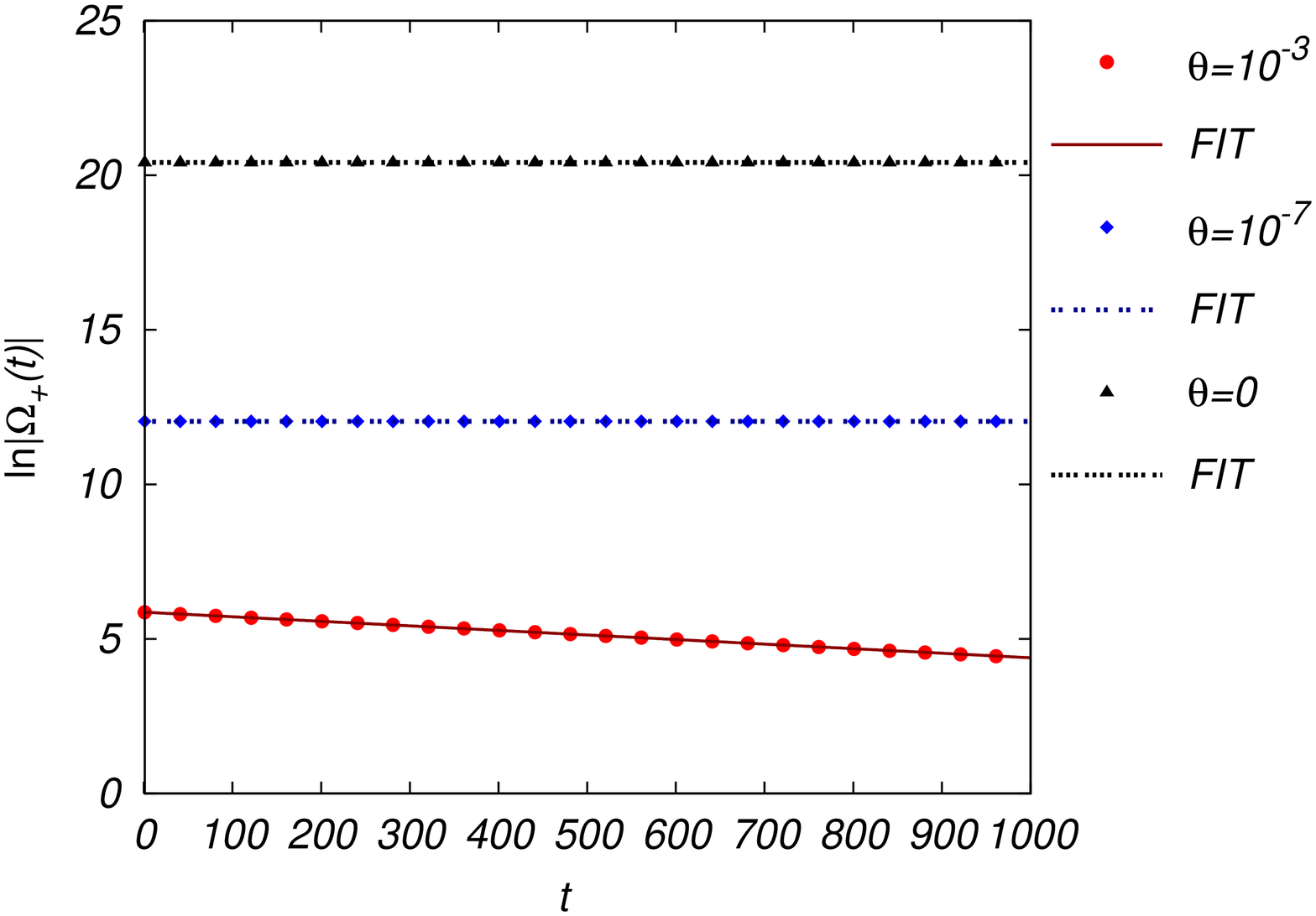}
\includegraphics[width=0.35\textwidth, angle=270]{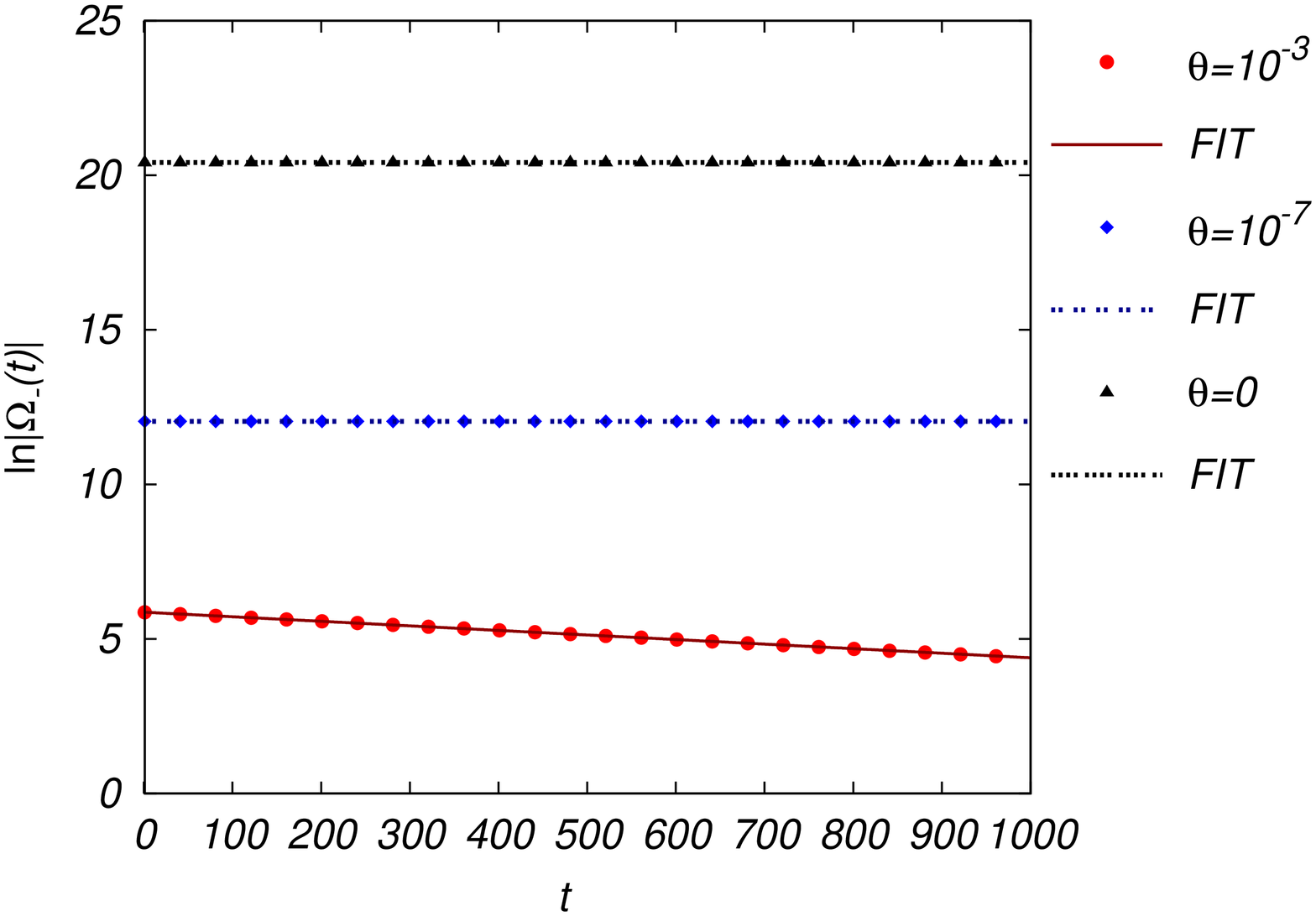}
\caption{Spatially averaged Schwinger functions involving the vector parts of the fermion propagator, Eq.~(\ref{spativec}), for $N=3$ and various values of $\theta$,  fitted according to Eq.~(\ref{fitomega}).  $\Omega_\pm(t)$ are positive definite, and thus  describe deconfined fermions.}
\label{testsigmav}
\end{center}
\end{figure}

\section{Concluding Remarks}
\label{conclusions}

In this article we have studied the dynamical generation of masses and confinement in Maxwell-Chern-Simons QED$_3$. Two species of fermions $m_+$ and $m_-$ have been considered within the four-component spinor formalism. These species are non-degenerate in mass. The origin of these physical masses is two-fold: On the one hand, there is a parity-preserving contribution $m_e$ coming from dynamical chiral symmetry breaking and, on the other hand, there is the Chern-Simons-induced Haldane mass term $m_o$.

In Ref.~\cite{BCRR} it was noticed that in the parity-invariant version of QED$_3$, assuming $N<N_c(0)$ and taking into account feed-back effects between the vacuum polarization and dynamical masses, the two fermion species are degenerate in mass and confined. In the $1/N$ approximation, however, this is not the case. Chiral symmetry for both species is broken as long as $N<N_c(0)$, but their charges get completely screened, regardless of whether the species remain in the chirally symmetric or asymmetric phase of the theory. Here we have considered parity-violating effects through the inclusion of the CS term in the Lagrangian. We observe that as $\theta$ increases, one of the two species becomes light and the other one becomes heavy. 

There exists a critical value $\theta_c$ at which the mass of the light species exhibits a discontinuity on $\theta$ and becomes negative, with the same magnitude as its heavy cousin. It is precisely at this value where chiral symmetry is restored~\cite{KondMarPRL,KondMarPRD}. The value $\theta_c$ depends on the number of fermion families in the model and is modified by  vacuum polarization effects for the photon. It follows a critical behavior similar to the chiral symmetry restoration pattern of ordinary unquenched QED$_3$, dominated by the value of $N_c(0)$. This explains the fact that zero physical masses can occur, in the $(N,\theta)$-plane of the model, along the segment $(N\ge N_c(0),\theta=0)$, provided, of course, the {\em infrared collusion} advocated in Ref.~\cite{BCRR} takes place. Throughout the entire $(N,\theta)$-plane, however, both species of fermions deconfine, as it was inferred from the local and global properties of the dynamically generated fermion propagator. This special feature of the $1/N$ approximation of dynamical masses without confinement, however, needs a deeper understanding of the effect of the CS term, because in this truncation, the above mentioned feed-back effects between dynamical masses and vacuum polarization are being neglected~\cite{BCRR}. This work is currently in progress. A natural extension is the inclusion of a finite chemical potential and/or temperature -- respective studies are also in progress. 

\begin{acknowledgements}
We acknowledge valuable discussions with A. Bashir, C.D. Roberts and P.C. Tandy. Support has been received from SNI, CIC and CONACyT grants through projects 4.22, 82230 and 50744-F, respectively. 
\end{acknowledgements}

\end{document}